\documentclass[12pt]{article}
\usepackage{natbib,float,amsmath,algorithm,algorithmic,doublespace}
\bibpunct[,]{(}{)}{;}{a}{}{,}
\begin{document}
\title{Errors drive the evolution of biological signalling 
to costly codes}
\author{Gonzalo G. de Polavieja \\
Dept. Zoology, Downing St.\\  University of Cambridge, CB2 3EJ, UK.\\
{\rm gg234@cus.cam.ac.uk}}
\date{}
\maketitle
\newpage

\begin{abstract}
\textbf{
Reduction of costs in biological signalling seems an evolutionary advantage, 
but recent experiments have shown signalling codes shifted to 
signals of high cost with an underutilisation of low cost signals. 
Here I derive a theory for efficient signalling that includes both errors and costs as constraints and I  
show that errors   
in the efficient translation of biological states into signals can 
shift codes 
to higher costs, effectively performing a quality control. 
The statistical structure of signal usage is 
predicted to be of a generalised Boltzmann form that penalises 
signals that are costly and sensitive to errors.  
This predicted distribution of signal usage against signal cost has two main features: 
an exponential tail required for cost efficiency and an underutilisation 
of the low cost signals required to protect the signalling 
quality from the errors. These predictions are shown 
to correspond quantitatively to the experiments in which gathering 
signal statistics is feasible as in 
visual cortex neurons.} 

\end{abstract}


SHORT TITLE: Errors drive signalling to costly codes

\newpage

\section{Introduction}

Cells, groups of cells and multicellular organisms communicate their states using signals. 
The types of signals and 
encoding mechanisms used can be very different but, irrespectively of the mechanism, 
signal transmission should have a high efficiency within biological constraints. 
A universal constraint is the signalling cost.
Have biological signalling codes evolved to minimise cost? 
Cost reduction seems advantageous 
\cite[]{Baddeley1,Baddeley2,Laughlin,Stemmler,Levy,Softky} 
but signalling systems might be simultaneously optimal 
not only respect to cost but also to other constraints resulting 
in signalling codes very different to the cost efficient ones. 
A second universal constraint is communication errors. 
Here I consider the extension of information theory \cite[]{Shannon, Cover} to include errors and cost together as 
constraints of signalling systems 
and find the optimal signal usage under these constraints. 
For clarity of exposition and because the best data sets for statistical analysis 
are in neural signals, I will particularise the discussion to cell signalling and 
discuss the relevance of results to other signalling systems afterwards. 

Neurons provide an experimentally tractable case of cell signalling.
The experimental evidence in neurons is counterintuitive. Neurons codes can 
underutilise low cost signals. 
For neurons using different spike rates as signals, it has been found that low rates 
that take lesser metabolic cost to produce are typically underutilised \cite[]{Treves}. 
Similarly, neurons using spike bursts as signals underutilise the bursts 
of one spike that would take lesser production cost \cite[]{Balasubramanian2, Balasubramanian1}. 
Theories of cost efficiency cannot explain these experimental results. 
According to the theories of cost efficiency, signalling systems should maximise their 
capacity to represent different states given a cost constraint or maximise 
the ratio of this representational capacity and the cost \cite[]{Baddeley1,Levy}. The optimal 
distribution for these theories is an exponential decaying with signal cost.
In this way the most probable signals are those of lowest cost in clear 
contrast to the underutilisation of the low cost signals observed experimentally. 
For this reason I consider here the evolution 
of biological signalling codes towards efficiency of transmission 
within the biological constraints of both cost and errors. 

This paper is organised as follows. Section \ref{section:general} gives 
the theoretical framework and the general result of optimal signal usage 
when both costs and errors constrain the signalling system. 
To find this optimal signal usage, an iterative algorithm that can be 
easily implemented is given. 
Section \ref{section:experiments} shows that the optimal solutions found 
predict quantitatively the experimental results for signal usage in visual cortex neurons. 
Section \ref{section:limits} gives the conclusions and discusses the application to 
a variety of biological signalling systems including animal communication for which it is shown 
that cheaters would shift efficient codes to high cost.

\section{Theoretical treatment}
\label{section:general}

For signal transmission between a signaller and a receiver to work, 
the signaller must use encoding rules that correlate its signalling states 
$C= \{ c_{1}, c_{2}, ..., c_{N} \}$ with the signals $S=\{ s_{1},s_{2},...,s_{N} \}$.
For intercellular signalling, the signals $S$ can be different values of concentration 
of the same chemical, different mixtures of several chemicals, different time patterns 
(say, different frequencies of spike generation 
 or bursts of different sizes), different spatial patterns 
or even different patterns of activation of a group of cells. 
The cellular states $C$ are the internal variables 
representing the ideal signals without errors. Experimentally, identical stimulations 
of the cell will produce a distribution of signals were the peak is the ideal noiseless 
signal corresponding to the cellular state and the variance comes from the errors.

The correlation of 
states and signals is subject to the constraints imposed by cost and errors.
We characterise these errors with the error matrix of conditional probabilities
$Q_{kj} \equiv p( c_{k}|s_{j} )$, a matrix given by the 
probability that the signal $s_{j}$ 
comes from the state $c_{k}$. When there are no errors present each signal 
comes from a single state, and the error matrix \textbf{Q} is diagonal. 
When there are errors present, there are nonzero nondiagonal elements.
The costs can be in molecular machinery (a convenient parameter can be the number 
of ATP molecules), in transmission times (for example, bursts of many spikes take 
longer times to transmit than of fewer spikes) and in risks (for example by the use 
of chemicals that can be toxic). 
We can formally write the  costs of producing the signals as $\epsilon_{kj}$ with for example 
$ \epsilon_{12}$ the cost for the conversion of the first state into the second signal. 
As we are interested 
in the signal usage, we refer the costs to the signals as 
$\epsilon_{j}=\sum_{k} Q_{kj} \epsilon_{kj}$.  
We always label the signals in order of increasing cost, 
$\epsilon_{1} \leq \epsilon_{2} \leq ... \leq \epsilon_{N}$.

We also need to formalise the notion of correlation between the signaller's states and the signals 
in order to consider the consequences of cost and errors for this correlation. We require 
a general measure of correlation that is valid for any nonlinear dependencies, unlike 
correlation functions \cite[]{Li}, and that does not use a metric that measures correlation in an 
arbitrary manner. 
The averaged \textit{distance} between the actual joint distribution    
$p(c_{i},s_{j})$ and the distribution corresponding to complete decorrelation 
$p(c_{i},s_{j})_{decorr} \equiv  p(c_{i}) p(s_{j})$ gives such a general measure of correlation of 
the form
\begin{equation} 
I(C;S)=
\sum_{i,j}p(c_{i},s_{j}) 
\log \biggl( \frac{p(c_{i},s_{j})}{p(c_{i}) p(s_{j})} \biggr),
\label{eq:correlation}
\end{equation} 
that is zero for the completely decorrelated case and increases with increasing correlation. 
This is the standard measure of statistical correlation used in communication theory 
where it is known as \textit{mutual information} \cite[]{Shannon}. 
The mutual information $I$ takes care of the errors as a constraint as 
it decreases for an error matrix with larger non-diagonal elements. To see this, we can 
write its expression in (\ref{eq:correlation}) in terms of the error matrix \textbf{Q} by separating 
it into the signal variability and the signal uncertainty terms as $I (C;S)=H(S) - H(S|C)$ ,
with $H(S)=-\sum_{j} p(s_{j}) \log p(s_{j})$ and $H(S|C)=\sum_{j} p(s_{j}) \xi_{j}$ with
\begin{equation}
\xi_{j}=-\sum_{k} Q_{kj} \log P_{jk}
\label{eq:noise}
\end{equation}
a measure of the signal uncertainty for signal $s_{j}$
and $P_{jk} \equiv p(s_{j}|c_{k}) $ the probability that the state $c_k$ produces the signal $s_j$. 
We can express $P_{kj}$ 
in terms of \textbf{Q} using Bayes' theorem as 
$P_{jk}=(p(s_{j}) Q_{kj})/ (\sum_{i} p(s_{i}) Q_{ki})$.
With these relations we see that the mutual information can be written as the difference 
of a term $H(S)$ that measures the variability of the signal and a term $H(S|C)$ 
that measures the signal uncertainty as the variability of the signal that comes from the errors in \textbf{Q}. 
This second term $H(S|C)$ is the constraint given by the errors.

Using the mutual information $I$ as the measure of correlation between states and signals, that 
includes the constraint given by the errors together with the cost constraint, 
we can now formulate precisely our problem. With which frequencies $p(s_{i})$ should 
the signals $S$ be used to have a high mutual information  $I$ 
between states $C$ 
and signals $S$ given the errors \textbf{Q} and the average cost 
$E=\sum_{i} p(s_{i}) \epsilon_{i}$ as the biological constraints? 
To answer this question we use the method of Lagrange multipliers (see Appendix). 
The solution of the equations obtained by this method can be found 
using different numerical methods and we have chosen 
the one given in Algorithm \ref{alg:algo1} 
based on the Blahut-Arimoto algorithm \cite[]{Blahut,Arimoto}, 
commonly used in rate distortion theory \cite[]{Cover}, 
because it is particularly transparent as to the form of the solution.
\begin{algorithm}
\caption{Optimal signal usage with noise and cost constraints}
\label{alg:algo1}
\begin{algorithmic}
\STATE Initialise the signal usage to a random vector $\bf{p}^{1}$.

\FOR{ $t=1,2$,...until convergence}

\STATE

\begin{equation}
 P_{jk}^{t} =\frac{p^{t} (s_{j}) Q_{kj}}{\sum_{j} p^{t} (s_{j}) Q_{kj}}  
\label{eq:step1b}
\end{equation}
\begin{equation}
 p^{t+1} (s_{j})=\frac{  \exp-\left( \beta^{t} \epsilon_{j} 
-\sum_{k} Q_{kj} \log P_{jk}^{t} \right) }{
 \sum_{i} \exp-\left( \beta^{t} \epsilon_{i} 
-\sum_{k} Q_{ki} \log P_{ik}^{t} \right)} , 
\label{eq:step2b}
\end{equation}
where $\beta^{t}$ 
in (\ref{eq:step2b}) has to be 
evaluated for each $t$ from the cost constraint
\begin{equation}
\frac{ \sum_{j} \epsilon_{j}\exp-\left( \beta^{t} \epsilon_{j} 
-\sum_{k} Q_{jk} \log P_{jk}^{t} \right) }{
 \sum_{j} \exp-\left( \beta^{t} \epsilon_{j} 
-\sum_{k} Q_{jk} \log P_{jk}^{t} \right)} = E.
\label{eq:betaeq}
\end{equation}

\ENDFOR

\end{algorithmic}
\end{algorithm}
From Algorithm \ref{alg:algo1}, we obtain that the optimal signal 
usage taking errors and cost as constraints is of the form in (\ref{eq:step2b})
\begin{equation}
 \widehat{p}(s_{j})=\widehat{Z}^{-1}  \exp \left( -\widehat{\beta} \epsilon_{j} - \widehat{\xi}_{j}  \right), 
\label{eq:general}
\end{equation}
where the hat on $p$, $Z$, $\beta$ and $\xi$ is a reminder that their values are 
obtained using the iterative Algorithm \ref{alg:algo1}. The expression for 
$\xi$ is given in (\ref{eq:noise}) and $Z$ is the normalisation constant.
This solution has a number of interesting characteristics. 
Both signal cost, through the term $\beta \epsilon_{j}$, and the signal uncertainty from the 
errors $\xi_{j}$, penalise the usage of the signal $s_{j}$ in an exponential form.
With no errors present the signal usage is a decaying exponential with the signal cost $\epsilon$. 
And with no cost constraint the signal usage is an exponential against the signal uncertainty from 
the errors $\xi$. The distribution for the error-free case coincides with the one obtained in 
Statistical Mechanics where it is known as the Boltzmann distribution. 
We name the general distribution including the effect of the errors in (\ref{eq:general}) 
as a generalised Boltzmann 
distribution. To obtain the general relationship between statistical correlation $I$ and 
average cost $E$, substitute the distribution in (\ref{eq:general}) in the expression for $I$ in 
(\ref{eq:correlation}) to obtain $\widehat{I}=\widehat{\beta} E + \log \widehat{Z}$,
where the parameter $\widehat{\beta}$ given in (\ref{eq:betaeq}) and the normalisation 
constant $\widehat{Z}$ are nonlinear functions of the average cost $E$. 
This expression is the most general relationship between mutual 
information and cost for efficient signalling.

Given the error matrix \textbf{Q}, 
an average energy $E$ and signals costs 
$\bf {\epsilon}$, that can be obtained either experimentally or from theoretical models, 
Algorithm \ref{alg:algo1} gives the optimal signal usage that maximizes signal quality 
while maximizing cost-efficiency. We can advance some characteristics of the signal usage 
for optimal communication. In biological systems we expect that the errors produced with highest 
probability are those with the lowest amplitude. Two examples illustrate this point. 
Consider first a cell that translates some states into signals but that 
when it is in a nonsignalling state, spontaneously produces signals by error.  
The most probable signals to be produced by error are those of lowest amplitude
and therefore lowest cost. This is the case in neurons when different values of spike 
rates are used as different signals 
and spontaneous signalling, say following a Poisson distribution, 
produces the highest rates with very low probability. 
The signals of lower rate have then a higher signal uncertainty and according to 
expression in (\ref{eq:general}) are then underutilized.
As a second example consider animal communication. According to the present 
framework, cheaters that can produce low-cost signals enter as errors in the communication 
between healthy animals. 
These errors make the low-cost signals to have higher uncertainty and, as in the case of neuronal 
signalling, according to (\ref{eq:general}) the low-cost signals should be underutilized. 

\section{Comparison with experiments}
\label{section:experiments}

The signal usage of a small percentage of neurons, 
$16 \%$ in the case of neurons in the visual cortex area MT of macaques \cite[]{Treves}, 
can be explained with a theory of cost-efficient signalling \cite[]{Baddeley1,Levy}. 
To explain the signal usage for the totality of visual cortex 
neurons we use the formalism 
presented in the previous section that not only requires signal efficiency but signal quality. 
As in \cite[]{Baddeley2}, the present formalism assumes a maximum signal variability 
with an energy constraint, the novelty here is to require also signal quality by minimizing signal 
uncertainty. We also assume that the spike rates are the symbols 
that the visual cortex neurons use to communicate \cite[]{Baddeley1,Baddeley2,Treves} 
and that the costs of each symbol in ATP molecules can be taken 
to be linearly proportional to the rate value. As a simple model to the main contribution from noise, 
we assume spontaneous signalling when the cell should be in a nonsignalling state. 
This random spike production is modelled by a Poisson distribution, with 
the average number of spikes produced by error in an interval as the single parameter 
that distinguishes different cells. The optimal signal usage 
obtained from the Algorithm $1$ for this case can then be approximated as (see Appendix)
\begin{equation}  
 p(\rm{Rate})=Z^{-1} \exp \left( -\exp \left(-\rm{Rate}/ \alpha \right) 
- \beta \rm{Rate} \right), 
\label{eq:generalb}
\end{equation}
where $Z$ is the normalization constant. 
Cost efficiency is assured by the term $- \beta \rm{Rate}$ 
that penalizes signals by their cost. Signal quality is assured 
by the term $\exp \left(-\rm{Rate} / \alpha\right) $ that penalizes signals 
by their signal uncertainty, that increases with $\alpha$. 
The predictions 
made by the optimal signalling in (\ref{eq:generalb}) are: 
(a) For high rate values the term required for signal quality in (\ref{eq:generalb}) 
is negligible, so 
optimal signal usage reduces to 
an exponential decaying with rate, that is, a straight line in a logarithmic plot. (b) Low rate values are expected to be 
underutilized respect to the straight line in (a). 
Specifically, the difference between the straight line in (a) and the logarithm of the probability, 
$-\beta \rm{Rate} -\log(p)$ must be a decreasing exponential. 
We compare these predictions to the rate distributions of inferior 
temporal cortex neurons of two rhesus macaques 
responding to video scenes that have been recently reported \cite[]{Treves}. 
The experimental distribution of rates for two of the cells (labelled 
as $ba001-01$ and $ay102-02$ in \cite[]{Treves}) are given in Figure $1$ using a $400$ ms window.
As seen in Figure $1$, the two predictions correspond to the experimental data. 
Cost-efficiency is responsible for the signal usage at high rates and both cost-efficiency 
and signal quality for the signal usage at lower values of rate. Different neurons may have different 
values of the average cost and different noise properties but the signal usage 
seems to be adapted to the optimal values for each cell.

\section{Discussion}
\label{section:limits}

We have proposed an optimization principle of coding that takes into account both the noise and the cost 
associated with the coding. The outcome of this principle is the prediction of the signal usage for 
efficient signalling systems. The optimal signal usage for a communication system constrained by errors and cost 
has been shown to have a generalised Boltzmann form in equation (\ref{eq:general}) 
that penalises signals that are costly and that are sensitive to errors. 
Noisy signals with low amplitude and therefore low cost are responsible in 
the evolution of signalling systems towards efficiency 
for a shift of signalling codes to higher cost to minimize signal uncertainty.
For the simplest case of linear costs and low cost noisy signals, 
the two main features of this optimal signal usage are an exponential tail at 
high cost signals needed for cost efficiency and an underutilisation of the 
low cost signals required to protect the signal quality against errors while 
maintaining the cost efficiency. The predictions made by this optimal signal usage have been shown to 
correspond to the experimental measurements in visual cortex neurons. 

We have so far discussed cell signalling, but as we noticed already in the Introduction 
we have chosen this particular type of signalling for concreteness. The theoretical framework 
here proposed does not require knowledge of the underlying mechanisms of signalling. The theory only 
uses the notion of statistical correlation of states and signals without the need to 
make concrete how this correlation is physically established and without any description of the 
types of signals except for the costs and errors. 
This is enough to understand the optimal signal usage with cost and error constraints. 
For this reason, the results apply generally to biological communication and also to 
non-biological communication. Intracellular communication and machine-machine communication
are two possible domains of application.
Another important case is animal communication for which game-theoretical models have predicted 
that the evolutionary incentive to 
deceit is overcame increasing the cost of signals \cite[]{Zahavi1,Grafen,Bradbury}. 
These costly signals are called handicaps 
and make the communication reliable in the sense of being honest. 
A different perspective is gained from the formalism presented here. Cheaters enter in a communication as 
errors in the communication between healthy animals and, as they are only able to produce low cost 
signals, the signal uncertainty of the low cost signals is higher. According to the 
general result in (\ref{eq:general}) these low cost signals should be underutilized by healthy animals 
for efficient communication. 
This means that signal quality requires a shift to high cost signals, as we saw in the case of neurons. 
In this case, cost can be metabolic, times or risks. In this way we obtain a statement of the 
handicap principle based on optimal communication without using the theory of games.
Provided we know the communication symbols, their cost and error characteristics, the 
present formalism would give the optimal use of symbols according to signal quality and cost-efficiency. 
In general, a combination of both theories with competition elements and signal quality should be used.

It is interesting to discuss the limits of the theoretical framework. 
First, we have assumed that errors and cost are the only constraints 
of the communication system. Although these constraints are universal, particular systems might have 
extra constraints, that can be added straightforwardly to the present formalism. 
However, even in the presence of new constraints, the effect of the errors of the low cost signals would 
be to shift the signalling code to higher cost. 
Second, we have argued that in biological communication systems the errors that are produced with 
highest probability are those of the lowest amplitude and therefore of the lowest cost. 
For efficient signalling, we have seen that the consequence of the noise of low cost signals is to 
shift siganls to a higher cost code. However, it is possible to have a more sophisticated noise structure that 
can affect the high cost signals. For example, 
processing of the signals at the receiver cell might fail more frequently for the 
most complex incoming signals, typically those with highest cost. In this case, there should be an 
extra penalisation of the high cost signals and the decay of the 
distribution would be faster than exponential. 
There is partial experimental evidence for this type of code in \cite[]{Baddeley2} 
(see their Figure 4(f,g)). In any case, the general result of the therey presented here is the 
generalized Boltzmann form in (\ref{eq:general}), that holds for any efficient signalling 
as it makes no assumptions about the noise or cost properties.

\section*{Acknowledgements}
Dennis Bray, William Bialek, Fabrizio Gabbiani, 
John Hopfield, Rufus Johstone, 
and Amotz Zahavi are acknowledged for fruitful 
discussions. I am especially indebted to Simon Laughlin for many discussions 
and critical comments on the manuscript. 
I am also very grateful to Vijay Balasubramanian and Michael J. Berry 
for discussing their independent results on metabolic efficiency prior to 
publication. I am thankful to Stephano Panzeri for sending me the data for Figure $1$ 
and for discussing the results in \cite[]{Treves}. This research has been supported by a 
Wellcome Trust Fellowship in Mathematical Biology. 

\newpage
\section*{Apppendix}

The optimization principle proposed consists in maximizing the mutual information 
subject to a cost constraint $E=\sum_{i} p(s_{i}) \epsilon_{i}$, 
where $E$ is the value of the average cost, $\{ \epsilon_{i} \}$ 
are the costs of the different signals and $\{ p_{i} \}$ the different probabilities of 
using the signals. Formally, using the method of Lagrange multipliers, 
we can write this optimization principle as
\begin{equation}
\max_{p(s_{i})} \left( I -\beta \left( \sum_{j} p(s_{j}) \epsilon_{j} - E \right) 
-\lambda \left( \sum_{j} p(s_{j}) -1 \right) \right),
\end{equation}
where the mutual information is given by 
\begin{equation} 
I=H(S)-H(S|C), 
\end{equation}  
with the entropy of the signal $H(S)$ given by 
\begin{equation}  
H(S)=-\sum_{j} p(s_{j}) \log p(s_{j}) 
\end{equation}  
and the entropy of the errors or noise $H(S|C)=-\sum_{j,k} p(s_{j}, c_{k}) \log p(s_{j} | c_{k})$ can be written as
\begin{equation}
H(S|C)=-\sum_{j} p(s_{j}) \sum_{k} Q_{kj} \log P_{jk}.
\end{equation} 
The matrix \textbf{Q} has elements $Q_{kj}=p(c_{k}|s_{j})$ 
given by the probability that the signal $s_j$ comes from the state $c_{k}$. 
The matrix $\textbf{P}$ has elements $P_{jk}=p(s_{j}|c_{k})$ given by the 
probability that a state $c_{k}$ produces the signal $s_{j}$, 
that can be written in terms of the probability of finding a signal $p(s_{j})$ 
and the matrix \textbf{Q} using Bayes' theorem as 
$P_{jk}=(p (s_{j}) Q_{kj})/(\sum_{i} p (s_{i}) Q_{ki})$. 
The optimization principle of coding includes both the errors 
through the error matrix \textbf{Q} (or \textbf{P}) and the costs associated with the coding.
The general solution to this optimization principle is numerical. 
Before discussing this general numerical solution, we consider two particular cases that are analytical.

\textit{All signals with same noise}. 
In this case the entropy of the noise reduces to a constant independent of the probabilities 
of using different signals, $H(S|C)=\alpha$. 
The optimization principle gives a result independent of the value of $\alpha$, 
with the probability of using a signal as an exponetial decreasing with cost (a Boltzmann distribution)
\begin{equation} 
p(s_{j})=\frac{\exp (-\beta \epsilon_{j} )}
{\sum_{i} \exp(-\beta \epsilon_{i})},
\end{equation}
with the parameter $\beta$ given by the average cost $E$ as
\begin{equation}
\frac{\sum_{i} \exp(-\beta \epsilon_{i}) \epsilon_{i}}
{\sum_{i} \exp(-\beta \epsilon_{i})}=E.
\end{equation}
For the case in which the cost is an average time $T=\int_{0}^{\infty}
d \tau p(\tau) \tau$, the Boltzmann distribution reduces to the Poisson distribution
\begin{equation}
p(\tau)=T^{-1} \exp(-T^{-1} \tau).
\end{equation}
\textit{A simple noise structure}. 
As a toy analytical model of the results presented in this paper, 
consider a simple case of three signals in which the first 
two require the same cost, $\epsilon_{1}=\epsilon_{2}$, and the third one a higher cost $\epsilon_{3} > \epsilon_{1}$ 
and with 
the noise matrix elements 
$p(c_{1}|s_{1})=p(c_{2}|s_{2})=1-\rho$, 
$p(c_{2}|s_{1})=p(c_{1}|s_{2})=\rho$ and 
$p(c_{3}|s_{3})=1$. This toy model has two noisy signals with lower cost and  
a higher cost signal with no noise.
The noise entropy for this case has the form
\begin{equation}
H(S|C)=
\left(p(s_{1})+p(s_{2}) \right) \xi .
\end{equation}
with $\xi = -\rho \log \rho -(1-\rho) \log (1-\rho)$.
The optimization principle for this example gives the probabilities
\begin{align}
p(s_{1,2}) &=p(m_{1,2})=Z^{-1} \exp ( -\beta \epsilon_{1} -\xi )   \\\     
p(s_{3}) &=p(m_{3})=Z^{-1} \exp (-\beta \epsilon_{3})   ,
\end{align} 
with $Z= 2 \exp (-\beta \epsilon_{1} -\xi)+ \exp(-\beta \epsilon_{3} ) $ 
the normalisation constant and $\beta$ given by the value of the average energy
$2 p(s_{1}) \epsilon_{1} +p(s_{3}) \epsilon_{3}=E$. The first two signals deviate from the 
Boltzmann form and are underutilized thus preserving signal quality.

\textit{General solution}. Differentiating the constrained mutual information and equating to zero gives 
the probability of using a signal of the form
\begin{equation}
 p(s_{j})=\frac{  \exp-\left( \beta \epsilon_{j} 
-\sum_{k} Q_{kj} \log P_{jk} \right) }{
 \sum_{j} \exp-\left( \beta \epsilon_{j} 
-\sum_{i} Q_{ki} \log P_{ik} \right)} , 
\end{equation}
but  $P_{jk} =(p (s_{j}) Q_{kj})/(\sum_{i} p (s_{i}) Q_{ki})$ 
also depends on $p(s_{j})$. This creates a nontrivial self-referential problem. 
However, the optimization of the mutual information 
respect to the probability of using a signal $p(s_{i})$ 
can be written as a double maximization (see Lemma 13.8.1 in \cite{Cover}), 
that is, the maximization of the mutual information,
\begin{equation}
\max_{p(s_{i})} \left( \sum_{j,k} p(s_{j}) Q_{kj} \log \frac{p(s_{j}) Q_{kj}}
{p(s_{j}) \sum_{m} p(s_{m}) Q_{km}} \right),
\end{equation}
can be written as the double maximization
\begin{equation}
\max_{P_{ij}} \left( \max_{p(s_{j})} \left( \sum_{j,k} p(s_{j}) Q_{kj} \log \frac{P_{jk}}{p(s_{j})} \right) \right).
\end{equation}
This double maximization suggests the possibility of an alternating maximization algorithm. 
Csiszar and Tusnady \cite[]{Csiszar} 
have shown that an alternating maximization algorithm for this problem converges to the required maximum. 
The algorithm starts with a guess of an optimal $p(s_{i})$ and with that calculates 
the conditional probability $P_{ij}$. This conditional probability is then used to recalculate 
a better guess to the optimal $p(s_{i})$ and the procedure is continued until convergence. 
This algorithm, including in our case the cost constraint, is given as Algorithm $1$ in the main text. 

For the case of a neuron, we would ideally include in Algorithm $1$ the experimentally measured values of the 
noise matrix \textbf{Q}, the signal costs $\{ \epsilon_{i} \}$ and the average cost $E$. 
As these values are not available from experiments at present, 
we consider the simplest models for both cost and noise. 
We consider a simple model of cost linearly proportional to the number of spikes in the time interval o finterest $T$, 
$\epsilon_{i} \propto i $ and the noise to be Poisson spontaneous signalling, 
$p(s_{i}|c_{0})=(\nu T)^{i} \exp (-\nu T)/i!$, 
with $\nu$ the frequency of spontaneous signalling and $T$ again the time interval of interest. For low $\nu$ 
this last expression can be approximated by an exponential, $p(s_{i} | c_{0}) \propto \exp (- \gamma i) $.  
Inserting these two approximations into Algorithm $1$, we find a good correspondence between theory and experiments 
for all cortex neurons tested, allowing 
for a different amount of spontaneous signalling parametrized by $\gamma$ and for a different average cost $E$ for 
each neuron. 
To obtain a simple analytical expression common for all cortex neurons, we fit the numerical data, or directly the 
experimental data, with the functional form suggested by the theory, 
$p(Rate; \alpha,\beta) \propto \exp( - \xi(Rate; \alpha) -\beta Rate)$ to find 
$\xi(Rate;\alpha) \propto \exp(-Rate/ \alpha)$ with different values of the parameter $\alpha$ depending on the 
noise of the particular neuron.

\newpage

FIGURE CAPTIONS

$    $

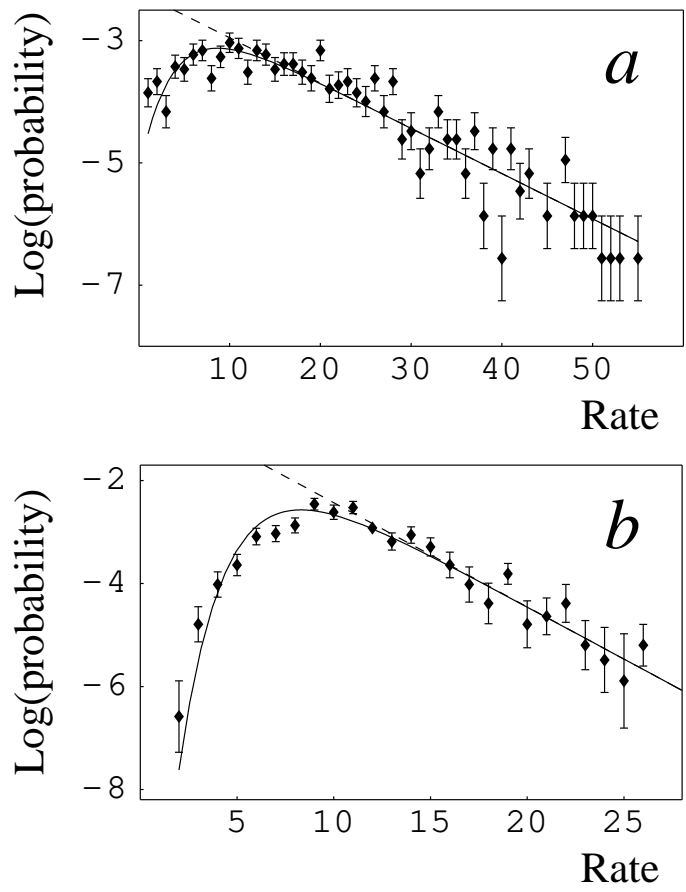
\textbf{Figure 1}. The probability distribution of rate usage for visual 
cortex neurons follows the optimal distribution in equation (\ref{eq:generalb}) (solid line) 
with the predicted exponential tail (dashed line) for high rates and the underutilisation at low costs. 
The exponential tail makes visual cortex neurons 
cost efficient and the underutilisation of the low cost signals protects their 
signal quality against errors while remaining cost efficient.
The errors are responsible for a shift to higher cost signals, with a maximum 
at a rate of value of $10$ spikes in the 400 ms window 
instead of at a rate of $1$ spike if there were no errors present.
The experimental data have been taken from the two visual cortex neurons 
labelled as (a) $ba001-01$ and (b) $ay102-02$ in \cite[]{Treves}.

$    $

\newpage
\bibliographystyle{/usr/users/gonzalo/gonzalo/texts/plainnat}
\bibliography{efficient.bib}
\newpage
\end{document}